\documentclass[10pt, conference]{IEEEtran}

\usepackage{cite}
\usepackage{amsmath,amssymb,amsfonts}
\usepackage{algorithmic}
\usepackage{graphicx}
\usepackage{textcomp}
\usepackage{xcolor}
\usepackage{soul}

\def\BibTeX{{\rm B\kern-.05em{\sc i\kern-.025em b}\kern-.08em
    T\kern-.1667em\lower.7ex\hbox{E}\kern-.125emX}}
\usepackage{hyperref}

\begin{document}

\newcounter{observation}
\newcommand{\observation}[1]{\refstepcounter{observation}
	\begin{center}
		\framebox{
			\begin{minipage}{0.93\columnwidth}
				{\bf Observation \arabic{observation}:} \textit{#1}
			\end{minipage}
		}
	\end{center}
}

\title{On the Nature of Code Cloning \\ in Open-Source Java Projects}

\author{\IEEEauthorblockN{Yaroslav Golubev,\IEEEauthorrefmark{1} \hspace{0.2em} Timofey Bryksin\IEEEauthorrefmark{1}\IEEEauthorrefmark{2}}
\IEEEauthorblockA{\IEEEauthorrefmark{1}JetBrains Research, Saint Petersburg, Russia, \hspace{0.2em}\IEEEauthorrefmark{2}Saint Petersburg State University, Saint Petersburg, Russia \\
\{yaroslav.golubev, timofey.bryksin\}@jetbrains.com}}

\maketitle

\begin{abstract}

Code cloning plays a very important role in open-source software engineering. The presence of clones within a project may indicate a need for refactoring, and clones between projects are even more interesting, since code migration takes place and violations are possible. But \textit{how} is code being copied? How prevalent is the process and on what level does it happen?

In this general study, we attempt to shed some light on these questions by searching for clones in a large dataset of over 23 thousand Java projects on the level of both files and methods, and by studying the code fragments themselves and their clone pairs. We study the size and the age of code fragments, the prevalence of their clones, relationships between exact and non-exact clones, as well as between method-level and file-level clones. We also discover and describe various anomalies in the code clones that were detected in the dataset.

Our research shows that the copying occurs all through the years of the Java code existence and that method-level copying is much more prevalent than file-level copying, with only 35.4\% of methods having no clones at all. Additionally, some of the discovered anomalies can be useful for future large-scale cloning research as they can be used for removing auto-generated code.

\end{abstract}

\section{Introduction}
The term \textit{code clones} refers to similar fragments of code that appear in different places. A lot of research has focused on identifying, studying, and combating code clones, because they are prevalent in modern day open-source software~\cite{lopes2017dejavu} 
and can bring about a lot of negative effects.

If cloning occurs within a single project, such clone pairs are called \textit{intra-project} and may indicate the necessity to refactor the project. Code clones may complicate the maintenance of the project, for example, because of bug propagation and functionality duplication, and the information about intra-project code clones helps the developers to fix the bugs more efficiently~\cite{chatterji2011measuring, chatterji2013effects}. \textit{Inter-project} clones (clones between different systems) can be of interest for discovering similar mobile applications~\cite{android} or determining potential candidates for libraries~\cite{libraries}. Another quite important application of detecting such clones is their possible relation to plagiarism and licensing violations~\cite{mathur2012empirical, golubev2020study}, a crucial topic in software engineering.

Code clones can occur at different levels of \textit{granularity}: statements, functions, classes, files, and even entire subsystems and systems. All such cases can be interesting for specific purposes: a coarse-grained view of code cloning can be used for discovering non-explicit forks between open-source systems, whereas fine-grained code clones can be employed for studying the migration of code snippets, especially given the popularity of services like StackOverflow~\cite{yang2017stack, romansky2018sourcerer}.

Code snippet clones are also traditionally divided into four \textit{types}~\cite{roy2007survey}. \textit{Exact clones} (Type-1) represent identical code fragments disregarding comments, whitespaces, and layouts. \textit{Renamed clones} (Type-2) also allow a difference in identifiers, types, and literals. These very precise clone types are of interest for discovering recognizable copying and pasting and possible licensing violations. On the other hand, \textit{Near-miss clones} (Type-3) allow more significant changes, like adding, removing, or moving specific statements, and \textit{Semantic  clones} (Type-4) provide similar functionality but differ syntactically. 

In this paper, we study the nature of code clones in a large corpus of open-source Java projects. The aim of the paper is two-fold: (1) to study how code clones of various granularities and types co-exist in open-source Java code and (2) discover anomalous code fragments and clones and study their impact on clone detection. We compile a dataset of more than 23 thousand Java projects from GitHub and search it for code clones on various levels using the SourcererCC clone detection tool~\cite{SourcererCC}: we detect near-miss clones of different size on the level of methods, as well as exact clones on the level of both methods and files. 

We study the size of the tokenized methods, their age, how many clones they have, and discuss some important anomalous examples. We also filter out clone pairs within one project, within one author, within forks and discovered mirrors to only consider clones from \textit{unconnected} projects that might constitute possible borrowings of code and study their differences from the standpoint of types and granularities. We analyze the temporal difference between clones, and for exact file-level clones we analyze their hashes, file names, and paths.

Our important findings include:

\begin{enumerate}
    \item Only 35.4\% of methods in our dataset have no clones at all. 31.9\% more only have clones within their project, and the remaining 32.7\% have clones in other projects.
    \item Methods with anomalously large number of clones very often come from auto-generated files. Interestingly, the vast majority ($>$95\%) of their clones are near-miss, not exact. The clones between auto-generated files also often have the same time difference between them and can be detected as anomalous spikes in the graph of time difference between clones.
    \item 31.5\% of unconnected exact file-level clones have different hashes due to small differences in whitespaces and comments, which indicates the limitations of using hashes to look for exact clones.
    \item Only 2.3\% of all exact unconnected method-level  clone  pairs  come  from  exact  unconnected  file-level clones. The rest come from different files, indicating the importance of considering method-level clones when conducting research.
\end{enumerate}

These findings can be used to better understand the nature of code cloning in open-source Java projects, and also to estimate the impact of anomalous and auto-generated files on the results of clone detection for more careful research in the future.

\section{Background}\label{background}

A lot of papers have focused on studying code clones in code systems and employed various approaches that targeted different types and granularities of clones. 
Earlier papers studied the coarse-grained possible reuse of code by studying the names of the files and repositories~\cite{mockus2007large}. The algorithm looked for directories that share several names of source files in them and where the fraction of such files is more than a certain threshold. 
The next, more thorough granularity involves searching for file-level clones. This level still remains one of the most popular one for conducting large-scale studies. For example, Ossher et al.~\cite{ossher2011file} studied over 13,000 Java projects and discovered file-level clones by comparing the following: firstly, MD5 hashes to find exact copies, then the fully-qualified names of the top-level types, and finally, name-based fingerprints of files to detect clones between files with more alterations. The findings of the paper showed that more than 10\% of all files are clones.

File-level clone detection is also very popular with \textit{clone detection tools} that are currently actively developed and perfected~\cite{ain2019recent}, which results in the fact that a lot of clone studies employ them. Lopes et al.~\cite{lopes2017dejavu} performed an exhaustive study on code clones in a dataset of several million projects in various programming languages. Once again, the authors employed several levels of clone detection: MD5 hashes of files, hashes of tokens in files, and, finally, near-miss clones. The authors concluded that unique files are a minority in their multi-language dataset.

However, file-level clones do not cover all possible cases of inter-project copying. Many works focus on function-level, or method-level, clone detection: in 2010, Roy et al.~\cite{roy2010near} discovered a lot of exact and near-miss function-level clones between systems. 
The reason for this is that open-source platforms like GitHub facilitate code reuse, and services like StackOverflow specifically deal with short code snippets. Yang et al.~\cite{yang2017stack} used SourcererCC~\cite{SourcererCC} on the level of methods to compare a large corpus of almost a million Python projects to almost two million snippets on StackOverflow. The results confirm the notion that there exists a migration between StackOverflow and GitHub. 
A similar study was later conducted by Romansky et al.~\cite{romansky2018sourcerer}, who researched the licensing violations between StackOverflow snippets and Python projects on GitHub. The authors also used SourcererCC on the level of methods and discovered that code migration co-exists with the constant relicensing of the code, which showcases another application of such fine-grained clone detection --- tracking license violations of the borrowed code.

A very interesting work was conducted recently by Gharehyazie et al.~\cite{gharehyazie2019cross}, who also studied the nature of code clones in Java. The authors used Deckard~\cite{jiang2007deckard}, an AST-based clone detection tool, to find clones among 8.5 thousand Java projects. The authors considered the difference between intra-project and inter-project clones, the influence of domains and developers' experience on the process of cloning, and studied the flow of clones in the software ecosystem. While their work inspired us, in our work, we focus on the methods themselves, different granularity and types of code clones, as well as on the temporal difference between them.

Finally, in our previous work, we used large-scale method-level clone detection to discover potential licensing violations in Java code on GitHub~\cite{golubev2020study}. We used SourcererCC on the dataset of more than 23 thousand Java projects and discovered that as much as 9.4\% of methods in the dataset could have been copied with potential licensing violations. The present work is the direct continuation of this research: while studying the clones that occur in the dataset, we discovered a lot of interesting observations and wanted to elaborate on them by adding the file-level clones into the picture and by manually evaluating specific cloned fragments.

\section{Methodology}\label{methodology}

In this paper, we used the dataset of projects and method-level code clones from our license violation study~\cite{golubev2020study}, so a more detailed description of data gathering can be found there. Here, we will briefly describe the entire pipeline.

\subsection{Dataset}

To study the nature of code clones in open-source Java projects, we compiled a large dataset based on the Public Git Archive (PGA)~\cite{pga}. PGA was compiled as a collection of all GitHub repositories with at least 50 stars as of early 2018. Since we were interested in the code's age, which requires full version control system history, we cloned all such repositories on June 1st, 2019. At the moment of the download, some repositories were already deleted or made private, some moved and pointed to the same projects, and some no longer had Java code in them. Such repositories were excluded, and the final dataset consisted of 23,378 Java projects. The full list is available online.\footnote{Dataset: \url{https://doi.org/10.5281/zenodo.3608211}}
We also used GitHub API to gather a list of forks for each project and identify the forks in our dataset.

\subsection{Clone detection}

The next step of the pipeline consisted in detecting clones in the collected corpus. All clone detection was carried out using SourcererCC~\cite{SourcererCC}, a clone detection tool that has proven itself to be precise and scalable. We had considered several different tools, but SourcererCC showed the best results at a reasonable time for such a large data corpus.
Clone detection was conducted at two different levels of granularity supported by the tool.

\subsubsection{Methods} 

The main interest of our study lied in the small-scale fragment-level copying of code. For that reason, firstly, we used SourcererCC in the \textit{methods} mode.\footnote{SourcererCC itself refers to this mode as \textit{blocks mode}, and refers to methods as \textit{blocks}. In this paper, we use the term \textit{method} for consistency.} We detected code clones with several parameter configurations as follows. The main search was conducted with the 
minimal token length threshold of 19 tokens to filter out trivial and universal code fragments and the similarity threshold of 75\%. In addition to this, we also conducted several searches with lower similarity thresholds and higher token length threshold: this allowed us to discover larger and less similar clones. The results of these searches were merged into one with the main search. More details about this technique and the exact parameters can be found in our previous work~\cite{golubev2021multi}.
Lastly, we ran a separate search with a similarity threshold of 100\% to discover \textit{exact} method-level clones in our dataset.

\subsubsection{Files}

As for the level of \textit{files}, we focused on only finding exact copies of the files and detected clones with the same lower token length threshold of 19 and the similarity threshold of 100\%, discovering exact file-level clones. The reason for this is that we were interested in the relation between exact method-level clones and exact file-level clones.

\subsection{Analysis}

Finally, we used the obtained code fragments and the information about clones within them to perform the analysis.

\subsubsection{Methods}
\label{sec:method_blocks}

We studied the size of the obtained methods in tokens and looked at some anomalously large examples. We also studied the methods from the standpoint of what clones they have: no clones at all, clones only within the same project, within \textit{connected} projects, or within \textit{unconnected} projects. In our work, we define \textit{connected} projects as:

\begin{itemize}
    \item projects within one user/organization;
    \item projects that are somehow connected by forking (direct fork, fork of a fork, common parent, etc.);
    \item manually discovered mirrors (indirect forks).
\end{itemize}

The rest of our analysis focused on \textit{unconnected} clone pairs (\textit{i.e.}, clones that were found between \textit{unconnected} projects), since they may constitute borrowings that demonstrate the actual propagation of code between projects.

For all methods that have unconnected code clones, we determined their last time of modification. We ran the \texttt{git blame} command on their original files and got the last date of modification for each line that constitute each method. We then aggregated this data for a method and defined the time of the last modification of a method as the most frequent last time of modification of its individual lines. More details and a demonstrating example can be found in our previous work~\cite{golubev2020study}.
We compared the time of the last modification of different methods and discovered the oldest ones.

\subsubsection{Method-level clone pairs}

In the obtained method-level clones, we firstly studied the distribution of the amount of clones among the methods and manually evaluated 50 random methods with an anomalously large number of code clones to discover what these methods are.

In the next step, we utilized the gathered exact clones. We analyzed the amount of exact clones, and for each method, we also analyzed the ratio of exact clones to all clones.

Finally, we analyzed unconnected method-level clone pairs from the standpoint of time difference between clones' last time of modification, detected outliers, and manually evaluated a random sample of 50 of these outliers to understand their nature and what projects they come from.

\subsubsection{File-level clone pairs}

Within the detected exact file-level clones, we firstly looked at them and studied what kind of files these are. We also studied these files to see what kind of clones they have, similarly to methods in Section~\ref{sec:method_blocks} (no clones, same project, connected, and unconnected projects).

Secondly, we analyzed their hashes to estimate how well the method of comparing hashes can find such clones. We manually investigated 50 random exact file-level clones that have different hashes and investigated the reasons. 

Next, we analyzed the similarity of file names and relative paths within their projects to quantitatively measure whether such files are \textit{utility} files~\cite{gharehyazie2019cross} --- files with general functionality that are copied without moving or changing. 

Lastly, we estimated how many exact method-level clones come from exact file-level clones.

\section{Results}\label{results}

Let us now present the main findings of our study.

\subsection{Methods}

The tokenization process on the level of methods revealed 38,617,427 different methods, of which 11,762,703 passed the token length threshold of 19 tokens and were therefore searched for clones.

\begin{figure}[h]
  \centering
  \includegraphics[width=3.3in]{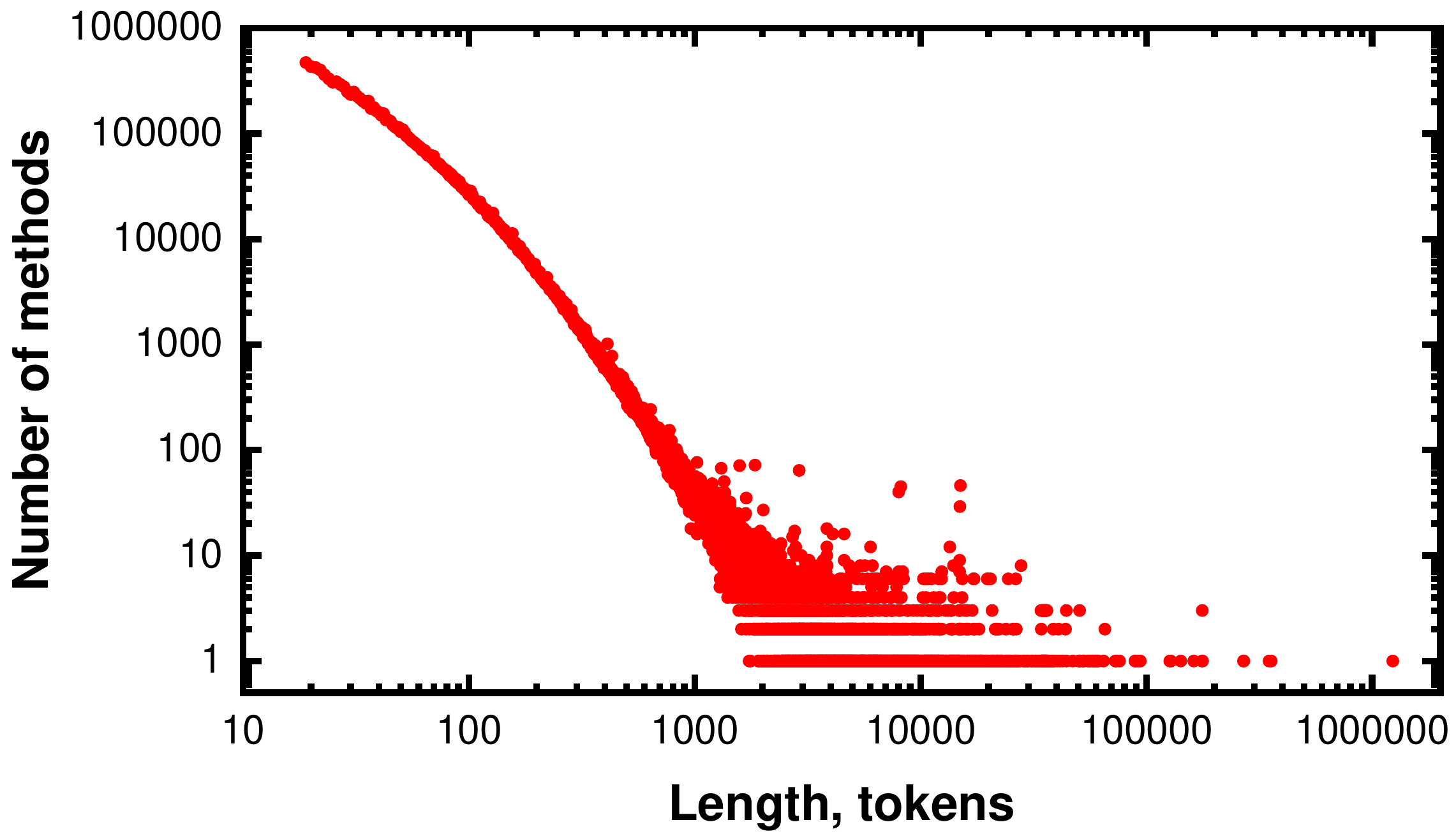}
  \caption{Methods that pass the token length threshold by their size in tokens.}
  \label{size}
\end{figure}

Figure~\ref{size} presents the distribution of these 11 million methods by their size in tokens. It can be seen that the dependence is stable up until the methods of approximately 1000 tokens, of which there are less than a hundred for each size. It can also be seen that there are anomalous methods more than 100,000 tokens in size. One method\footnote{A method with more than a million tokens: \url{https://github.com/fraunhoferfokus/Fuzzino/blob/75d20f05db8dfbfe035a6df0b9eb7c88703886fc/src/main/java/de/fraunhofer/fokus/fuzzing/fuzzino/heuristics/generators/string/data/SpecialUnicodeBomStrings.java}} even has more than a million tokens, although, interestingly, only 12 of these tokens are unique: these are special Unicode strings. Such methods can be unwanted in various applications of clone detection and introduce noise to the dataset.

\observation{Large datasets of raw data may contain anomalously large methods. Researchers should be aware of such files and the fact that they may often contain unusual and atypical code, and may apply some upper token length thresholds, if necessary.}

\begin{figure}[h]
  \centering
  \includegraphics[height=1.9in]{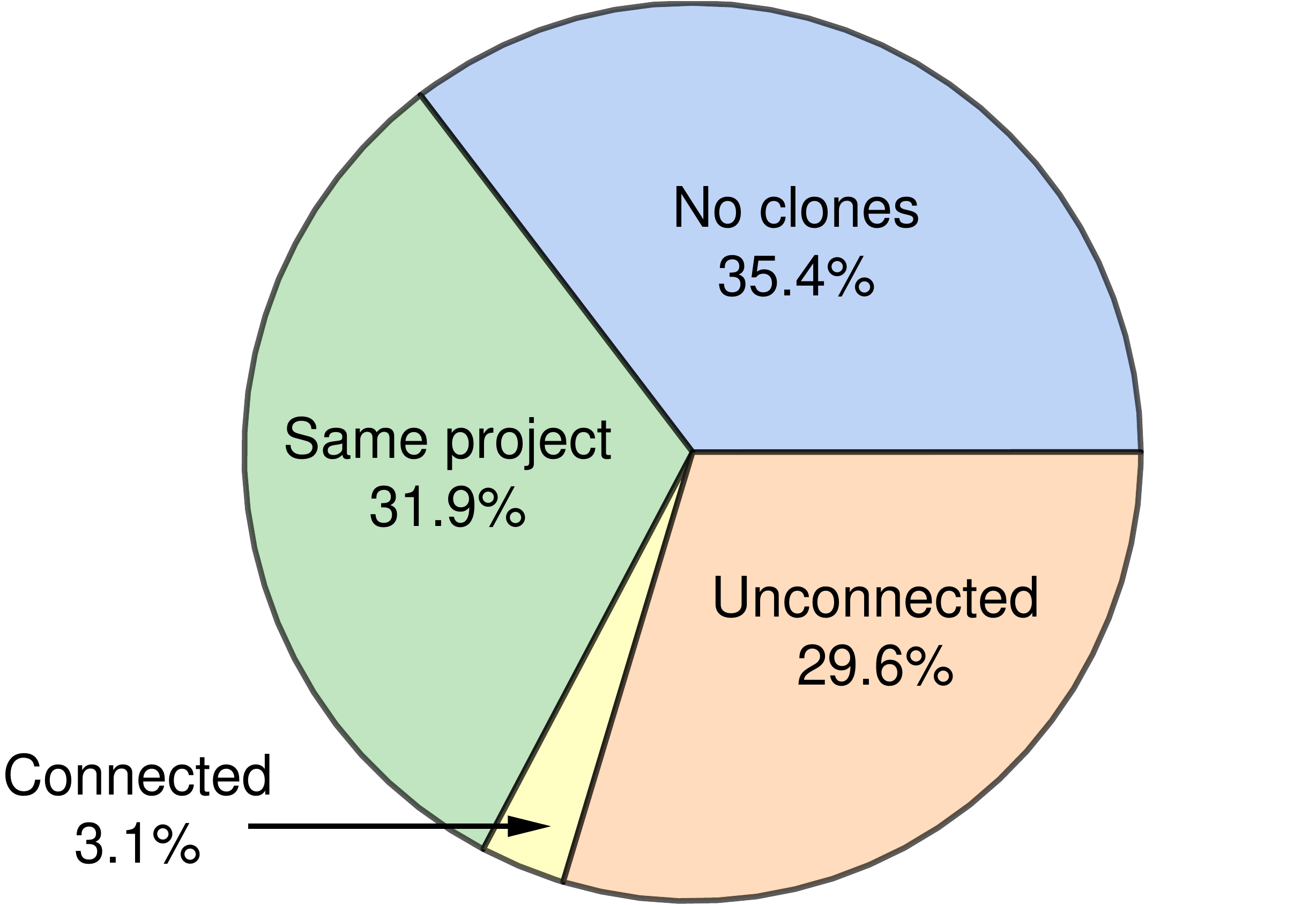}
  \caption{Methods that pass the token length threshold by their domain of clones.}
  \label{domain}
\end{figure}

After a non-exact method-level search was conducted for these methods, we divided the methods into four domains of cloning: methods that have no clones, methods that have clones only within their own project, methods that have clones in connected projects (same user or organization, forks, or mirrors), and methods that have clones in unconnected projects. Their distribution is presented in Figure~\ref{domain}.

It can be seen that there are approximately equal amount of methods that have no clones at all, have only clones within a project, and that have clones in other projects. 

\observation{Code cloning is very prevalent on the level of methods in our dataset, with only 35.4\% of them being fully unique.}

\begin{figure}[h]
  \centering
  \includegraphics[width=3.3in]{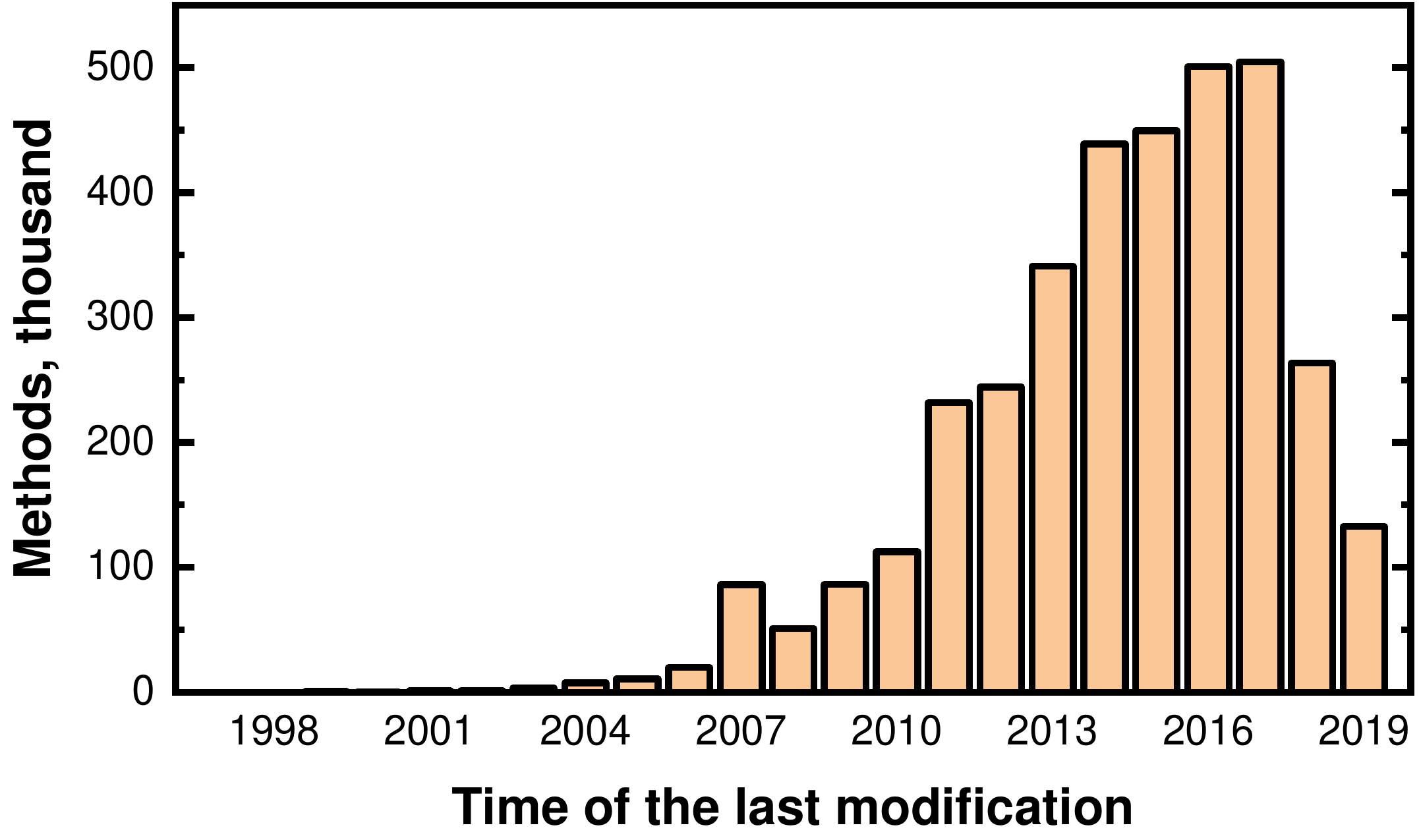}
  \caption{Methods with unconnected clones by their time of last modification.}
  \label{time}
\end{figure}

Since we are only interested in comparing the last time of modification for clone pairs that can constitute possible borrowings, we calculated this time using \texttt{git blame} only for methods that have clones in unconnected projects. The distribution of these times is presented in Figure~\ref{time}.

Considering the fact that the dataset was compiled in the early 2018 and cloned in the middle of 2019, it can be seen that the most of the largest parts of methods' bodies were written in 2016-2017, indicating a constant modification of code and its constant addition. The oldest methods, two of them, come from January 4th, 1997, from the file that implements Diff for files.\footnote{File that contains methods changed in 1997: \url{https://github.com/IanDarwin/javasrc/blame/2fabfa1c660bd717428968a257b48df2d6a988f1/main/src/main/java/textproc/Diff.java}} The commit message reads: \textit{Total rewrite: convert entire program from C to Java}.

\subsection{Method-level clone pairs}

A total of 1,163,989,420 clone pairs were detected on the level of methods. Of them, 560,656,419 (48.2\%) were inter-project.
As was mentioned above, to study possible borrowings, we excluded the pairs from the same user/organization, from forks and from discovered mirrors. To counter the possible presence of undiscovered mirrors, as well as to only consider clone pairs that could be real borrowings, we additionally excluded pairs with methods with the same date of the last modification. All the following results are presented for the remaining clone pairs.

\begin{figure}[h]
  \centering
  \includegraphics[width=3.3in]{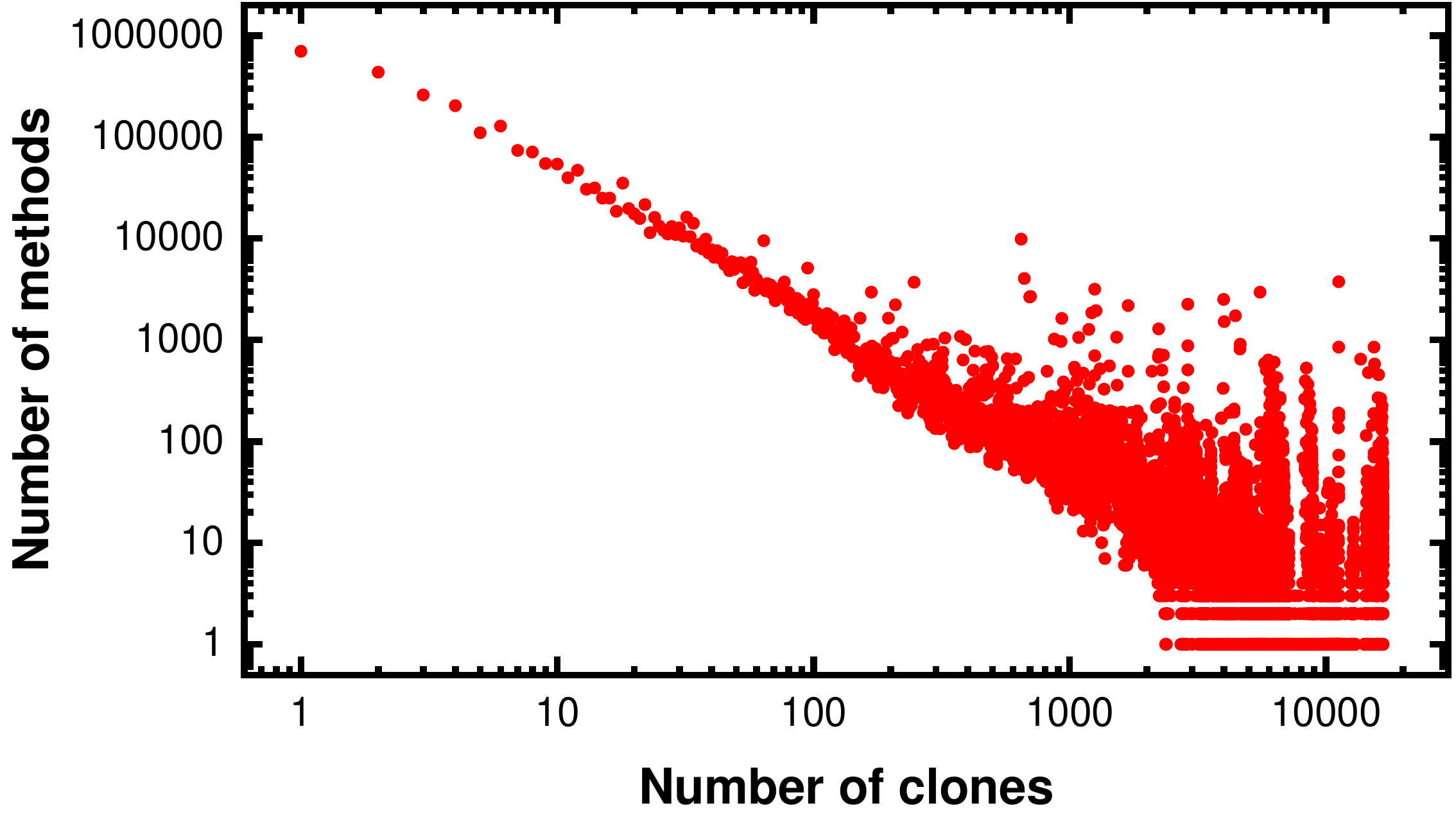}
  \caption{Methods that have at least one unconnected clone by the amount of unconnected clones they have.}
  \label{clones}
\end{figure}

Firstly, we wanted to study the distribution of clone pairs among methods. The distribution of methods by their amount of clones is presented in Figure~\ref{clones}.

It can be seen that this distribution is also very extreme. More than half of methods only have 5 clones or less, the distribution is more or less the same up until about a hundred clones, after which it gets chaotic, with some methods having up to 17,000 unconnected clones. We manually evaluated 50 random methods with more than 10,000 clones, and discovered only two kinds of methods. The majority are methods from auto-generated files (explicitly marked in the header), and several other methods are \textit{equals()} methods that are also often auto-generated. Interestingly, all of these most cloned auto-generated methods have only 3-4\% exact clones. 

\observation{Methods with anomalously large number of near-miss clones very often come from auto-generated files. However, only about 3-4\% of their clones are exact. Researchers should be mindful of the importance of near-miss clones in this regard and may use an upper threshold for the number of clones if they wish to filter out auto-generated files.}

Of course, ideally, researchers should filter out auto-generated files preemptively, based on the comments at the top of the file, however, the proposed heuristic allows to filter the data after the search, as well as possibly catch cases of auto-generated methods in otherwise usual files.

\begin{figure}[h]
  \centering
  \includegraphics[width=3.3in]{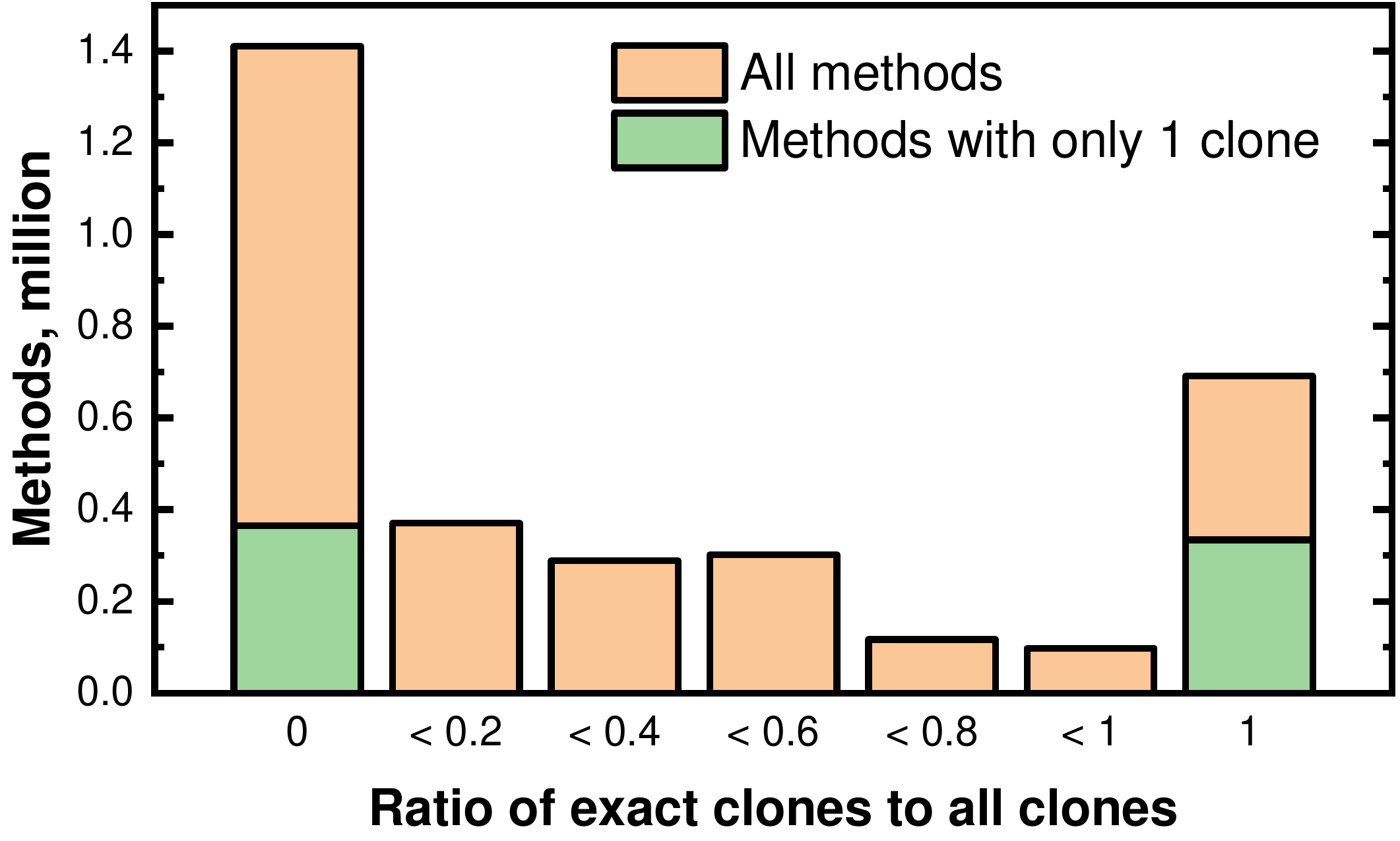}
  \caption{Methods by the ratio of exact unconnected clones to all unconnected clones. Green area at the bottom of columns indicates methods that only have a single clone.}
  \label{ratio}
\end{figure}

Next, we looked at the discovered exact, Type-1, clones. In total, from these unconnected clone pairs, exact clone pairs represent 11.7\%, which is quite a lot. In absolute numbers, this is more than 50 million exact clone pairs.
However, the sheer percent of exact clones does not give us the full picture of their distribution. Figure~\ref{ratio} presents the distribution of methods by their ratio of exact clones to all clones. 

43\% of methods that have at least one clone, have no exact clones at all. Because of the prevalence of methods with small number of clones, we also visualized the fraction of methods that only have one non-exact clone.
35.9\% more of methods have both exact and non-exact clones in various proportions.
Finally, 21.1\% of cloned methods only have exact clones. In order for this data not to look misleading, we also highlighted methods that only have one clone (which is exact). However, even without them, 10.9\% of methods with any clones have more than one clone, all of which are exact. Even though these methods might have other clones within their own project, author, or fork, their high percent still indicates that this is a specific type of methods. 

\begin{figure}[h]
  \centering
  \includegraphics[width=3.3in]{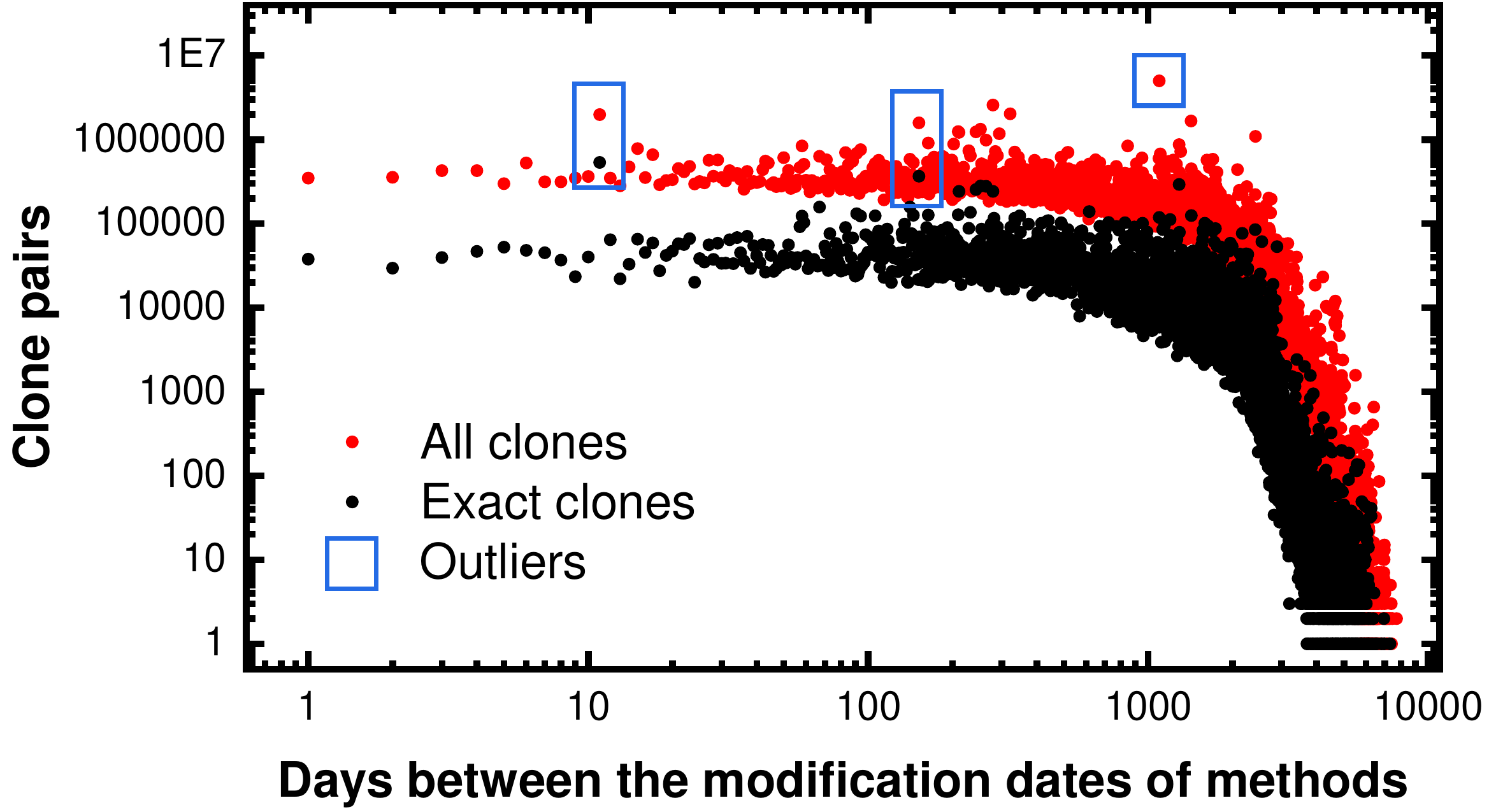}
  \caption{Clone pairs between unconnected methods by the time difference between the methods, in days. Blue rectangles indicate some of the outliers on the graph.}
  \label{time_dif}
\end{figure}

Having the information about the last time of modification of methods, it is also possible to consider the time difference between methods in clone pairs. The distribution of clone pairs by the time difference between the methods is presented in Figure~\ref{time_dif}.
The time difference spans through all the years in our dataset. Similar methods can be separated by days, weeks, months, years, and in rare cases even decades. 

In general, the distribution is more or less equal until we get to 5-6 years of difference, when there becomes significantly fewer clone pairs. Mainly, this has to do with the fact that the majority of code is recent: Figure~\ref{time} shows that only 10.9\% of methods come from 2010 or earlier. It is possible that the style of coding also has an impact, this can be investigated in future on a modified balanced dataset.

Interestingly, the largest difference in time between methods occurs two times --- with two of the above-mentioned methods from 1997. That implementation of Diff was uploaded with slight modifications in 2018, making for a record difference of 21 years.

Another important feature of this distribution is the presence of visible spikes. Notice that this cannot be cause by indirect forks (when the history of a project is pushed to another repository), because with the saving of a history the time difference between corresponding clones would be 0. By looking at the projects where clone pairs come from, we discovered that clone pairs with anomalous time difference come from specific pairs of projects. We looked at 50 clone pairs with such time differences from these projects and discovered that the pairs that we have examined were all among sizeable auto-generated files. Indeed, it makes sense: such files are large and all the methods in them have the same date of the last modification. Therefore, projects with similar auto-generated files will produce a lot of method-level clones with the exact same time difference. 

\observation{Similar auto-generated files may produce an anomalously large number of clones with the same temporal difference. When working with the temporal aspect of code cloning and code evolution, researchers may remove such anomalies to filter the dataset.}

\subsection{File-level clone pairs}

\begin{figure}[t]
  \centering
  \includegraphics[height=2.1in]{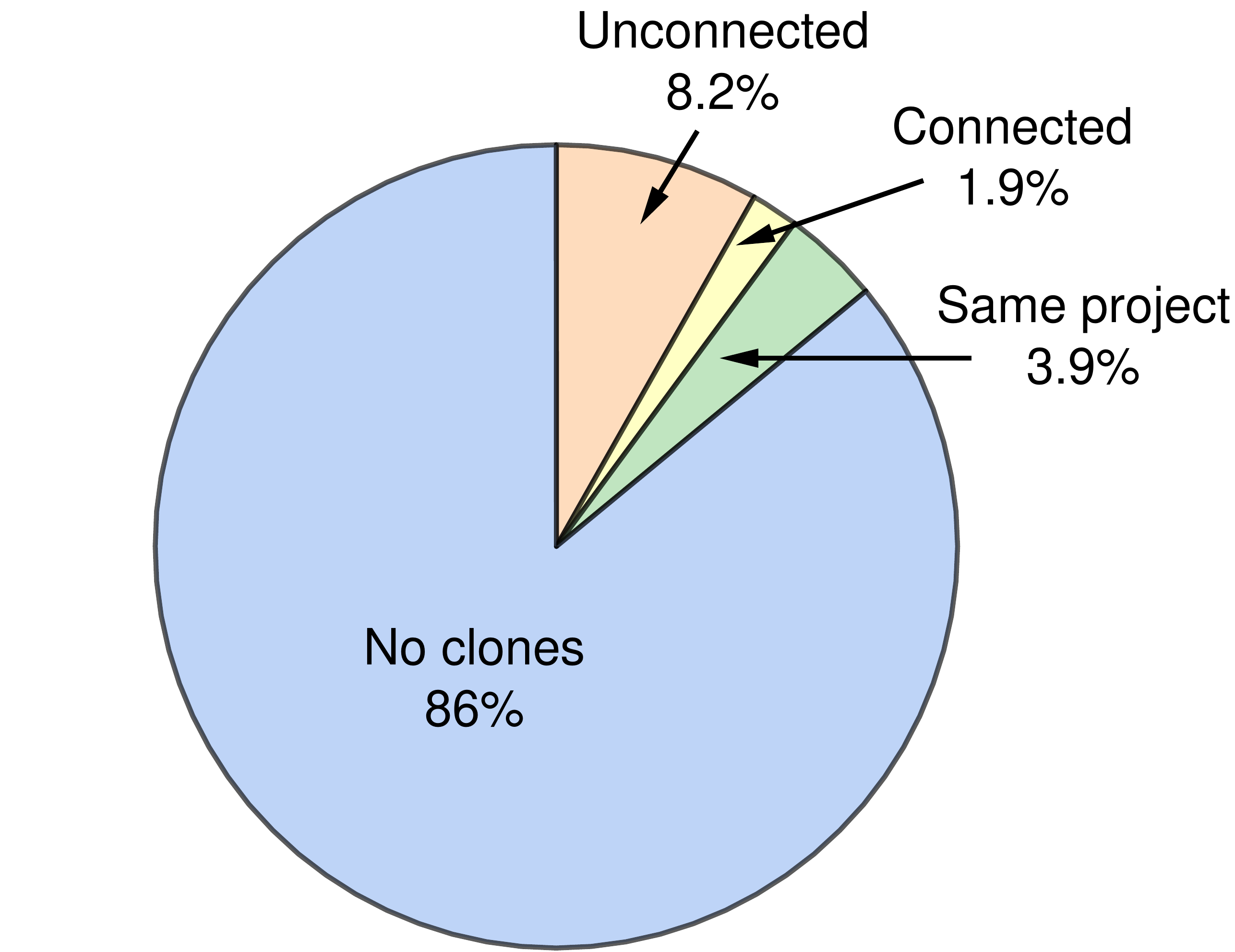}
  \caption{Files that pass the token length threshold by their domain of exact clones.}
  \label{domain_files}
\end{figure}

The tokenization on the level of files resulted in 4,145,534 bags of tokens for files, of which 3,922,290 passed the token length threshold of 19. Their distribution by size looks very similar to Figure~\ref{size}, with the same stable region up to approximately 1000 tokens (although with a spread out peak at 19--50 tokens) and a more chaotic region after. The largest file\footnote{The file with the most tokens: \url{https://github.com/cinchapi/concourse/blob/0321cad4c38a26c487c1511e26945eadea1aa11c/concourse-driver-java/src/main/java/com/cinchapi/concourse/thrift/ConcourseService.java}} has more than 800,000 lines of code and more than 1,500,000 tokens. 
The distribution of files by their exact clones domains is presented in Figure~\ref{domain_files}. 

The file-level search was conducted only for exact clones, therefore, a significantly larger percentage of files---86\%---has no such clones. Interestingly though, as much as 3.9\% of files have exact file-level clones in their own project. Exact method-level clones in the project are understandable, whereas exactly the same files within one repository indicate onto something unusual. This occurs in 1,950 different projects, although, as usual, the list is dominated by several dozens of projects with a lot of such cases. In the examples we looked at, this happened because of some projects storing different versions of applications in parallel directories, the same general-purpose files in several external libraries, as well as in projects that consist of collections of various programs, like programming tasks or proofs of concept.

SourcererCC also provides a hash for every tokenized file. We were surprised to discover that as much as 31.5\% of discovered unconnected exact clone pairs have different hashes. We have manually evaluated 50 such pairs and discovered differences in whitespaces and comments. Often, the difference is in the copyright of the license. Also, sometimes developers copy the file from another project but add some explanatory comments. 

\observation{31.5\% of discovered unconnected exact clone pairs have different hashes. Researchers who rely on hashes for discovering exact file-level clones should be aware of this, and our result can serve as an estimation as to how much exact clones may be lost this way.}

After that, we looked at the clones themselves. Unsurprisingly, we found that the vast majority of such files can be called utility files that present various universal functionality or belong to common libraries that are copied in their entirety. 

\begin{figure}[t]
  \centering
  \includegraphics[width=3.3in]{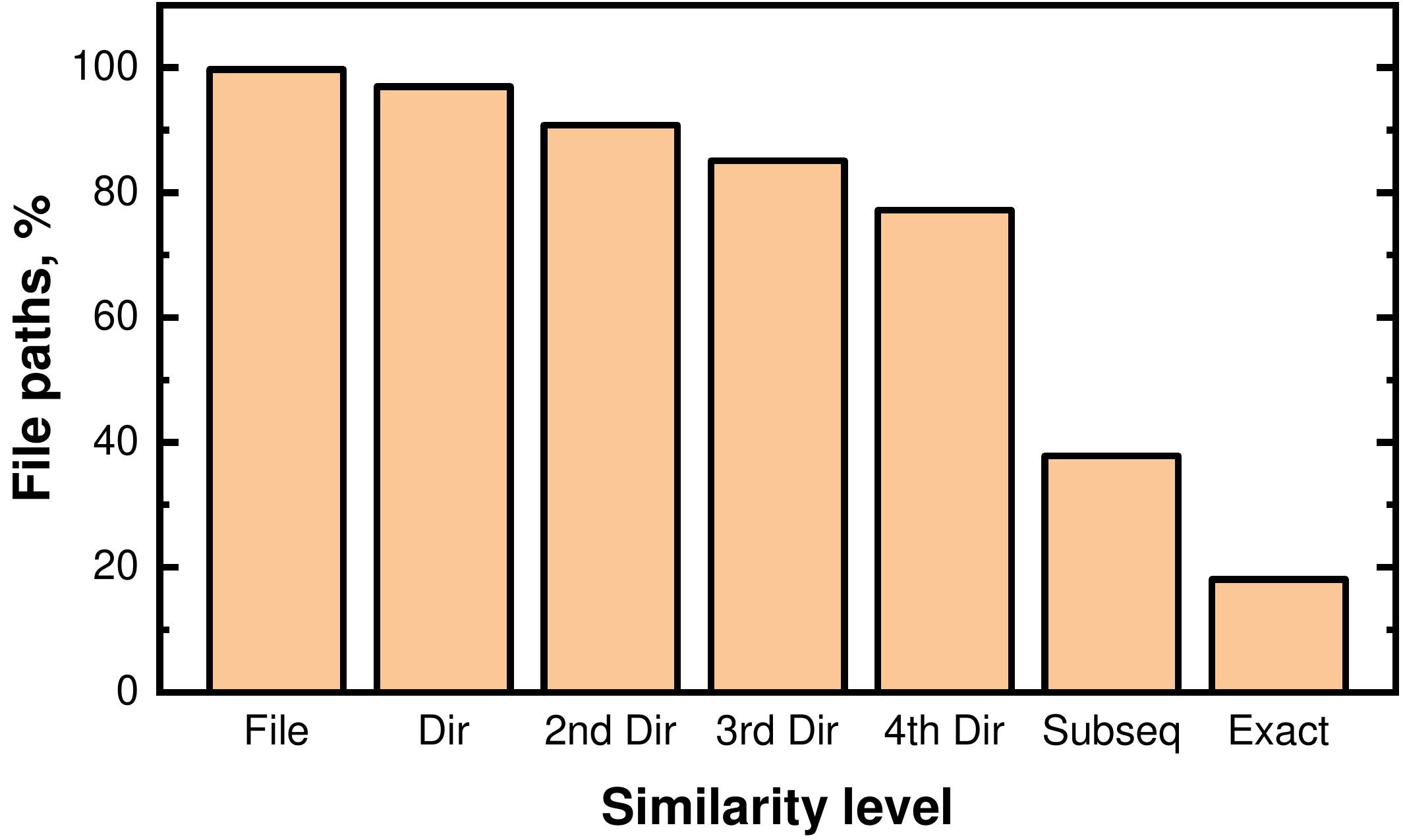}
  \caption{The percentage of file paths to exact file-level clone pairs with different levels of proximity. File --- same file name, Dirs --- same parent directory names at various levels, Subseq --- the path to one file is a subsequence of the path to another, Exact --- the paths are identical.}
  \label{names}
\end{figure}

To try and estimate this effect quantitatively, we measured the similarity of the files' names and relative paths. This comes from the observation that such auxiliary files are usually saved in very nested directories in Java. The results of this analysis are presented in Figure~\ref{names}.

Naturally, 99.7\% of exact file-level clones share their name. We looked at the cases where it is not so, in some cases, one of the files is named \textit{Test \#} or \textit{Sample \#}.
97\% of pairs also share the name of their directories, and this trend continues for the levels of parent directories. We checked up to the fourth directory (five levels, including the file), and 77.2\% of clone pairs share this much of the directory tree. To find out files that are copied with preserving the structure even more, we also calculated that 37.8\% of clone pairs have one of their relative paths be a subsequence of another (for example, \textit{/dir1/dir2/file.java} and \textit{/third-party/dir1/dir2/file.java}), and 18.1\% of clone pairs have their paths completely the same.

All this supports the idea that file-level clones very often represent popular utility files and are copied as a part of an entire directory or module. Another, much rarer, instance of clones includes cases where the files are copied to be used as a test for something.

Finally, to connect our granularities of clone detection, we calculated how many exact method-level clone pairs come from exact file-level pairs. It turned out that this is only 2.3\% of all unconnected method-level clone pairs. The other 97.7\% of exact method-level clone pairs come from files that have some differences. 

\observation{Only 2.3\% of all exact unconnected method-level clone pairs come from exact unconnected file-level clones. Researchers should be mindful of the importance of considering fragment-level code clones in the studies.}

\section{Threats to validity}\label{threats}

\textit{Internal}. In our work, we mainly focus on studying code clones that can represent potential borrowings of code. We not only filter out intra-project clones, we also consider the users and organization, forks, mirrors, and the difference in time between the code fragments. However, we still cannot realistically identify all possible borrowings. Nonetheless, we believe that the general nature of our study can be used for insights into the nature of code clones.

\textit{External}. Firstly, we only considered GitHub as a platform, and secondly, we used the repositories from Public Git Archive, meaning that they have more than 50 stars. It is possible that a larger sample of Java projects, perhaps less popular ones, can give a more thorough insight into the types of code clones. We also may not be sure that random samples of methods and clone pairs that we used for manual evaluation will generalize to all other cases. Still, our dataset is large and diverse, and therefore can be used for studying general trends. 

\textit{Construct}. SourcererCC, like all clone detection tools, detects some false positive clones, but we believe that its precision and recall discussed in previous works~\cite{SourcererCC, golubev2021multi} are appropriate for a large scale study like this one.

Overall, while the general nature of our study implies certain threats to validity, we believe that they do not invalidate the results.

\section{Conclusions}\label{conclusions}

In this study, we aimed to research the diversity of code clones in open-source Java projects, better understand their nature, and measure it quantitatively. In order to do that, we studied near-miss and exact method-level clones, as well as exact file-level clones.

In our dataset of 23,378 Java projects, 35.4\% of methods have no clones at all, 31.9\% only have clones within their projects, 3.1\% have clones within \textit{connected} projects (same author, forks, mirrors), and, finally, the remaining 29.6\% of methods have at least one clone in \textit{unconnected} projects. Half of such unconnected methods have 5 clones or less, with the most being about 16 thousand clones for a method --- these methods with anomalously large number of clones very often come from auto-generated files. Some clones between large auto-generated files can also be detected when working with the time difference between code clones.

Exact file-level clone pairs often concern universal, general-purpose files. We discovered that 99.7\% of such clone pairs share the name of the file, 77.2\% of clone pairs share their parent directories up to the fourth level, and 18.1\% of files share their entire relative paths. However, it is important to note that 31.5\% of these clones differ in hashes, which can be caused by a slight change in whitespaces, comments, or copyrights and should be remembered when using hashes for discovering clones. Also, only 2.3\% of exact method-level clones come from these exact file-level clones, indicating onto a prevalence of method-level code cloning in migration.

There are several possible directions for the future work in the field of large-scale clone detection. It is possible to use a larger dataset with less popular projects or repeat this study in another language and see if the discovered results generalize well to other languages. Also, several parts of this study can be explored more deeply, for example, the age of code. It would be interesting to study how many of the fragments were committed at a single point of time, and how many evolved and changed --- and how this affected their amount of clones and the nature of the clones that they have.

\bibliographystyle{ieeetran}
\bibliography{cites}

\end{document}